\documentclass[superscriptaddress,longbibliography,nofootinbib,twocolumn]{revtex4-2}
\usepackage[english]{babel}

\usepackage{graphicx} 
\usepackage[]{xcolor}

\usepackage{amssymb}
\usepackage{amsmath}
\usepackage{amsthm}
\usepackage{mathtools}

\usepackage{bm}
\usepackage{bbold} 
\usepackage{physics}
\usepackage[version=4]{mhchem}
\usepackage{siunitx}

\definecolor{LEI-blue}{cmyk}{1,.75,0,.35} 
\definecolor{LEI-orange}{cmyk}{0,.62,.97,0} 
\definecolor{niceblue}{rgb}{.1, .25, .8}

\usepackage[colorlinks=true, linkcolor=niceblue, citecolor=gray]{hyperref}

\newcommand{\topic}[1]{%
}

\DeclareMathOperator*{\argmin}{arg\,min}

\newcommand{\dens}{\boldsymbol{\rho}}
\newcommand{\pot}{\boldsymbol{\mu}}
\newcommand{\gs}{\text{GS}}

\newcommand{\EUT}{F} 
\newcommand{\tildeEUT}{\tilde{F}} 

\newcommand{\aQa}{\affiliation{%
    $\langle a Q a^L \rangle$ Applied Quantum Algorithms
    ---
    Lorentz Insitute for Theoretical physics \& Leiden Institute of Advanced Computer Science,
    Universiteit Leiden, The Netherlands}
}

\newcommand{\VU}{\affiliation{%
    Theoretical Chemistry, Vrije Universiteit Amsterdam, The Netherlands}
}
\newcommand{\cofirst}{
    \altaffiliation{These authors contributed equally to this work}
}

\begin{document}

\title{Learning Density Functionals from Noisy Quantum Data}

\author{Emiel Koridon}
\cofirst
\aQa
\VU

\author{Felix Frohnert}
\cofirst
\aQa

\author{Eric Prehn}
\aQa

\author{Evert van Nieuwenburg}
\aQa

\author{Jordi Tura}
\aQa

\author{Stefano Polla}
\email[e-mail: ]{polla@lorentz.leidenuniv.nl}
\aQa

\date{July 2023}

\begin{abstract}
    The search for useful applications of noisy intermediate-scale quantum (NISQ) devices in quantum simulation has been hindered by their intrinsic noise and the high costs associated with achieving high accuracy.
    A promising approach to finding utility despite these challenges involves using quantum devices to generate training data for classical machine learning (ML) models.
    In this study, we explore the use of noisy data generated by quantum algorithms in training an ML model to learn a density functional for the Fermi-Hubbard model.
    We benchmark various ML models against exact solutions, demonstrating that a neural-network ML model can successfully generalize from small datasets subject to noise typical of NISQ algorithms.
    The learning procedure can effectively filter out unbiased sampling noise, resulting in a trained model that outperforms any individual training data point.
    Conversely, when trained on data with expressibility and optimization error typical of the variational quantum eigensolver, the model replicates the biases present in the training data.
    The trained models can be applied to solving new problem instances in a Kohn-Sham-like density optimization scheme, benefiting from automatic differentiability and achieving reasonably accurate solutions on most problem instances.
    Our findings suggest a promising pathway for leveraging NISQ devices in practical quantum simulations, highlighting both the potential benefits and the challenges that need to be addressed for successful integration of quantum computing and ML techniques.
\end{abstract}

\maketitle

\begin{figure*}[t]
    \centering
    \includegraphics[width=1\linewidth]{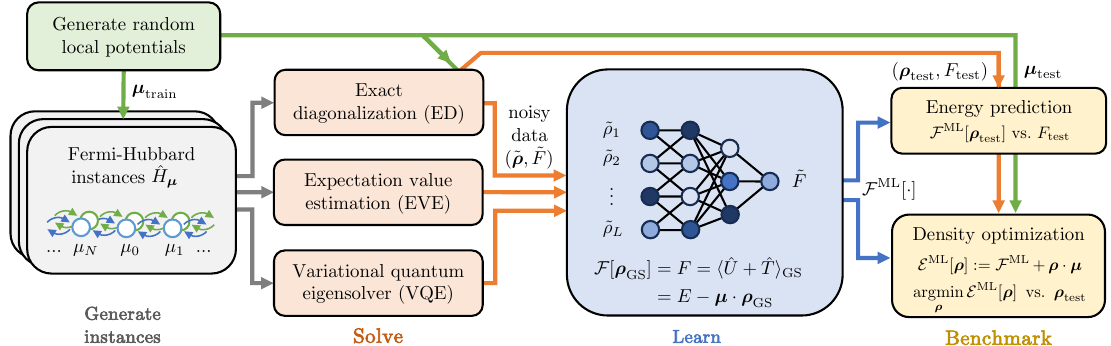}
    \caption{\textbf{Overview:} 
    Workflow for machine learning a DFT functional from noisy quantum data.
    The process begins with generating instances of the Hubbard model $\hat{H}_{\pot}$ from random local potentials $\pot_\text{train}$.
    For each of these instances, estimates of the ground state density $\tilde{\dens}$ and the potential-interaction energy $\tilde{\EUT}$ are obtained using one of three different methods: ED, EVE or VQE.
    The data extracted from EVE and VQE are noisy, representing two distinct types of errors --- handled separately in this study --- that are characteristic of NISQ quantum algorithms: sampling noise for EVE and expressibility and optimization errors for VQE.
    The pairs $(\tilde{\dens}, \tilde{\EUT})$, split in training and validation sets using $5$-fold cross validation, are used to train a machine learning model to learn the universal functional $\mathcal{F}$.
    This, in turn can be used to predict the ground state energy $E_\gs = F[\dens_\gs] + \pot \cdot \dens$, and to perform density optimization for new potentials (in a Kohn-Sham-like scheme).
    The ML models are benchmarked on a test set of new problem instances defined by random potentials $\pot_\text{test}$ and solved with exact diagonalization.
    The trained models are benchmarked on both of these tasks using a separate test set from ED to investigate the robustness to errors. 
    }
    \label{fig:overview}
\end{figure*}

\section{Introduction}

\topic{Quantum simulation, NISQ and noise}
Following the rapid development of quantum hardware, the last decade saw an explosion in the research on quantum algorithms for noisy intermediate-scale quantum (NISQ) devices~\cite{preskillQuantum2018}.
An important goal of this research is identifying avenues towards achieving useful quantum advantage: the application of quantum devices to relevant, classically-intractable problems.
A prominent example is 
the electronic structure problem, which is central to chemistry and material science, and has been a major focus for quantum algorithms due to its inherent quantum nature and significant scientific and commercial importance \cite{reiherELucidating2017, bauerQuantum2020, mcardleQuantum2020}.
Much of the recent research on quantum algorithms for this problem has centered on the variational quantum eigensolver (VQE), a NISQ-tailored algorithm designed to approximate ground states by optimizing a heuristic quantum ansatz~\cite{peruzzoVariational2014,cerezoVariational2021}.
However, several significant challenges hinder the achievement of practical solutions to the electronic structure problem. 
Accurately estimating energies and other properties from a state prepared on a quantum computer is made prohibitively expensive by sampling costs \cite{weckerProgress2015a}.
This is exacerbated by the necessity of applying error mitigation techniques to reduce bias due to circuit noise, which increases the sampling overhead \cite{caiQuantum2023}. 
Additionally, VQE introduces errors intrinsic to the algorithm, related to the expressibility of the ansatz and the complexity of the optimization landscape.
In practical electronic structure calculations, it is often necessary to solve for multiple system configurations, which increases the overall cost.

\topic{Density functional theory and learning functionals}
Density Functional Theory (DFT) is the workhorse of modern quantum chemistry and material science, used to investigate the electronic structure of many-body systems at a low computational cost \cite{burkeDFT2013}.
Central to DFT is the definition of a \emph{universal functional}, which maps the ground state electronic density to the corresponding kinetic and interaction energy for a given problem family \cite{hohenbergInhomogeneous1964,kohnSelfConsistent1965}.
Although exact universal functionals theoretically exist, their general explicit form is unknown, and their evaluation would necessarily be computationally intractable \cite{schuchComputational2009}.
Instead, DFT practitioners rely on a vast array of approximate functionals, developed over the decades through mathematical assumptions and physical intuition.
Recently, machine learning (ML) approaches have emerged as powerful tools for designing new DFT functionals  \cite{snyderFinding2012, liUnderstanding2016}.
In particular, deep-learning-based functionals have demonstrated remarkable performance on benchmark problems in chemistry \cite{grisafiTransferable2019,ryczkoDeep2019, kirkpatrickPushing2021}.
These models can be trained on mixed datasets generated through various approaches, including DFT based, on other functionals, expensive first-principles methods like coupled-cluster, and experimental data \cite{cohenChallenges2012, christensenSemiempirical2016}.
The synergies between DFT and quantum algorithms have also been explored by a few works, notably in the context of learning functionals on fault-tolerant quantum computers \cite{bakerDensity2020}, and in using quantum algorithms to supplement results from approximate DFT functionals \cite{sheridanEnhancing2024}.

\topic{classical ML with quantum data}
As quantum hardware continues to improve, the prospect of using quantum devices to generate training data that is inaccessible to classical methods is becoming increasingly realistic. 
Such data, derived from the quantum simulation of complex systems, could significantly enhance the capabilities of classical ML models \cite{huangPower2021, huangProvably2022, lewisImproved2024}.
In turn, these models could generalize from the limited, expensive-to-generate quantum data. 
Moreover, by leveraging the underlying structure, physically-motivated ML models could combine information from all the training data points, filtering out noise and leading to predictions that are more accurate than any individual training point.
While not much research has focused on practically training classical models with quantum-generated data, related concepts exist in the literature. 
In quantum science, ML is often applied to learning properties of a system from measurement outcomes, in tasks like phase classification \cite{dawidModern2023}.
Classical shadow tomography \cite{huangPredicting2020} constructs classical models that can predict many properties of a given quantum state from a limited set of measurements. 
Similarly, recent work demonstrates a technique to extract of classical surrogates from QML models trained on quantum devices  \cite{jerbiShadows2024}. 
Moreover, classical ML models have been employed for noise mitigation in quantum systems \cite{liaoMachine2023}.

\topic{in this paper} 
In this work, we explore the application of NISQ devices to generate noisy training data aimed at learning DFT functionals for the Fermi-Hubbard model, with the goal of generalizing and enhancing these datasets.
We focus on limited, noisy datasets of 1,000 data points, which, though small from a machine learning perspective, can already be expensive to generate on a quantum device.
We separately examine the effects of sampling noise typical of NISQ algorithms and the algorithmic errors associated with VQE, analyzing how these factors impact the learned functionals.
The models are benchmarked on both an energy prediction task and a Kohn-Sham-like density optimization task.

The remainder paper is structured as follows: 
Sec.~\ref{sec:background} summarizes the necessary background on DFT, the physical system and the learning task.
Sec.~\ref{sec:dataset} discusses the dataset generation process, including details on the considered quantum algorithms.
Sec.~\ref{sec:learning_functional} describes the machine learning model, training process, and compares various ML regression methods.
Sec.~\ref{sec:results} presents the benchmarking results of the trained ML models.
Finally, Sec.~\ref{sec:conclusion} discusses the findings and provides an outlook for future research.

\section{Background}\label{sec:background}

\subsection{Density Functional Theory}

Central to DFT are the Hohenberg–Kohn theorems \cite{hohenbergInhomogeneous1964}, which assert two fundamental principles: (1) the many-body ground state of a system of interacting electrons in an external potential is uniquely determined by its electron density, and (2) the ground state energy can be obtained variationally by minimizing a functional of the electronic density.

To formalize this and set the notation, consider a family of electronic Hamiltonians
\begin{equation}
    \hat{H} = \hat{U} + \hat{T} + \hat{V}
\end{equation}
where $\hat{U}$ represents a fixed electron-electron interaction, $\hat{T}$ is a fixed kinetic energy term, and $\hat{V}$ is a free external potential term. 
The expectation value of the potential $\langle \hat{V} \rangle$ is defined as a linear functional of the electronic density ${\dens}$.

The first Hohenberg-Kohn theorem ensures the existence of the \emph{universal functional}
\begin{equation} \label{eq:universal-functional}
    \mathcal{F}: {\dens}_{\gs} \mapsto \EUT := \bra{\psi_{\gs}}(\hat{U}+\hat{T})\ket{\psi_{\gs}}, 
\end{equation} 
which maps the electron density ${\dens}_{\gs}$ of the ground state $\ket{\psi_{\gs}}$ of any $\hat{H}$ to the respective expectation value of the kinetic and interaction energy $\EUT$.
The functional is termed \textit{universal} because it is independent of the external potential acting on the system.
The total ground state energy $E = \bra{\psi_{\gs}} \hat{H} \ket{\psi_{\gs}}$ can be reconstructed from the ground state density as 
\begin{equation} \label{eq:total-energy-functional}
    E = \mathcal{E}[{\dens}_{\gs}] := \mathcal{F}[{\dens}_{\gs}] + \bra{\psi_{\gs}}\hat{V}\ket{\psi_{\gs}},
\end{equation}
where $\bra{\psi_{\gs}}\hat{V}\ket{\psi_{\gs}}$ is a linear functional of ${\dens}_{\gs}$ by definition.
The second Hohenberg-Kohn theorem states that, for a given $\hat{V}$, the ground state energy and density of the corresponding $\hat{H}$ can be obtained by minimizing this functional: 
\begin{equation} \label{eq:second-hk-theorem}
    E = \min_{\dens} \mathcal{E}[\dens] 
    \,,\quad
    \dens_\gs = \argmin_{\dens} \mathcal{E}[\dens].
\end{equation}

While DFT approaches have been developed to investigate a wide variety of many-body systems, here we focus on a particular lattice model. 
In lattice models, electrons occupy discrete, localized sites, and the electron density $\dens$ is represented as a vector where each component $\rho_j$ denotes the occupation of the $j$-th site. 
The functional $\mathcal{F}[\dens]$ is the reformulated as a function of this vector, simplifying the computational treatment of the system and making it more amenable to a machine learning scheme. 

\subsection{The Fermi-Hubbard model}

In this work, we focus on the Fermi-Hubbard model, a paradigmatic lattice spin-$\frac{1}{2}$ Fermion model with on-site interaction.
We take this model on a one-dimensional ring of $L$ sites.
The Hamiltonian of this model,
\begin{equation}\label{eq:hubbard_ham}
    \hat{H}_{{\pot}} = \hat{T} + \hat{U} + \hat{V}_{{\pot}},
\end{equation}
comprises a kinetic energy term $\hat{T}$, ab interaction term $\hat{U}$, and a potential term $\hat{V}_{\pot}$. 
All of these can be expressed in terms of fermionic creation and annihilation operators, $\hat{c}_{j,\sigma}$ and $\hat{c}^\dagger_{j,\sigma}$.
From here on, $j \in \{1, ..., L\}$ denotes the site index, and $\sigma=\{\uparrow, \downarrow\}$ represents the electron spin.
The number operator is defined as $\hat{n}_{j, \sigma} = \hat{c}^\dagger_{j,\sigma} \hat{c}_{j, \sigma}$.
The kinetic term describes the hopping of electrons between adjacent sites and is given by:
\begin{equation}\label{eq:kinetic_term}
    \hat{T} := 
    - t \sum_j^L \sum_{\sigma \in \{\uparrow, \downarrow\}} (\hat{c}^\dagger_{j,\sigma} \hat{c}_{j+1, \sigma} + \hat{c}^\dagger_{j+1,\sigma} \hat{c}_{j, \sigma}),
\end{equation} 
where $t$ is the hopping parameter.
We choose without loss of generality $t=1$, thereby fixing the units of all energies.
We consider periodic boundary conditions, meaning ($\hat{c}_{L+1,\sigma} := \hat{c}_{1, \sigma}$).
The interaction term accounts for the on-site repulsion between electrons of opposite spins and is expressed as:
\begin{equation}\label{eq:interaction_term}
    \hat{U} := 
    u \sum_j^L \hat{n}_{j, \uparrow} \hat{n}_{j, \downarrow},
\end{equation}
where $u$ is the interaction strength. 
The potential $\hat{V}_{\pot}$ defines a specific instance of the Hubbard model within the broader family of models characterized by the chain length $L$ and the interaction strength $u/t$. 
The local chemical potential ${\pot} = (\mu_0, ..., \mu_L)$ introduces an additional term at each site to the Hamiltonian:
\begin{equation}\label{eq:potential_term}
    \hat{V}_{{\pot}} 
    = {\pot} \cdot \boldsymbol{n} 
    := \sum_{j}^{L} \mu_j (\hat{n}_{j, \uparrow} + \hat{n}_{j, \downarrow}).
\end{equation}

Each choice of ${\pot}$ defines a unique instance $\hat{H}_{{\pot}}$ within the model family, with ground state  $| \psi_{\gs} \rangle$, and the corresponding ground state energy $E$.
The elements of the electronic density vector $\dens_\gs$ are the expected number of particle at each site, summed over the two spin species:
\begin{equation}
    \rho_j = \expval{(\hat{n}_{j,\uparrow} + \hat{n}_{j,\downarrow})}{\psi_\gs}.
\end{equation}

The universal functional $\mathcal{F}$ is then a function mapping the vector ${\dens}_\gs$ to the scalar $\EUT = \bra{\psi_{\gs}}(\hat{U} + \hat{T})\ket{\psi_{\gs}}$; we call this the \emph{Hubbard functional}.
Note that the Hubbard functional depends on the value of the interaction strength $u/t$, the number of sites $L$ and the geometry of the system.

In our numerical study, we focus on a periodic Hubbard chain of $L=8$ sites; this system is small enough to allow effective exact diagonalization benchmarks. 
We choose an interaction strength of $u/t=4$, corresponding to the most challenging regime to simulate when scaling to larger systems \cite{leblancSolutions2015}.
Furthermore, we restrict all solutions to the quarter-filling ($L/4 = 2$ particles of each spin species) and total-spin singlet subspace, based on the symmetries of the Hubbard model.
A more general functional can in principle be obtained by stitching together functionals defined on different symmetry sectors.

\subsection{The learning task}\label{section:learning_task}

In recent years, data-driven approaches for learning approximate DFT functionals have emerged, where a machine learning model is trained to map electron densities to energies \cite{snyderFinding2012}.
For the Hubbard model, Nelson \emph{et al.} \cite{nelsonMachine2019} demonstrated a method for training a model to learn the functional $\mathcal{F}$ using 105,000 density-ground state energy pairs obtained via exact diagonalization. However, the reliance on exact diagonalization limits the feasible system size due to its high computational demands.
In this work, we consider the task of learning an approximation $\mathcal{F}^\text{ML}$ of the same functional, but from noisy data generated by a quantum algorithm.
Generating such data is expensive, with the cost rising with both the size of the dataset and the desired accuracy.
By teaining a model on a limited dataset, we aim to enable generalization and reduce noise, potentially leading to predictions with greater accuracy than any individual data point.

Our overall approach to learning a DFT functional from noisy quantum data is illustrated in Fig. \ref{fig:overview}: 
The process begins by initializing the Hubbard model with random values of ${\pot}$ drawn from a specified distribution (Sec.~\ref{sec:potentials-and-ED}).
The ground state problem is then solved for $1000$ potentials to obtain a dataset of noisy density-energy pairs $(\tilde{\dens}_{X}, \tildeEUT_{X})$, where $X$ indicates the method used to generate the data.
To represent the noise typical of NISQ algorithms, we consider and simulate two methods: expectation value estimation (EVE), which isolates the effect of sampling noise, and the variational quantum eigensolver (VQE), which isolates expressibility and optimization errors.
Exact diagonalization results are used for benchmarking, while the noisy data are employed to train a neural network-based regression model.
This approach allows us to explore how the inherent noise characteristic of quantum data from current near-term devices impacts the training of a machine-learned DFT functional.

\section{The dataset}\label{sec:dataset}

Training and testing data sets are constructed by solving configurations of the Hubbard problem with a random on-site potential ${\pot}$.
The same configurations are solved with each of the considered methods: exact diagonalization (ED), expectation value estimation (EVE), and variational quantum eigensolver (VQE).
The training set and the test set are generated using $1000$ independent random potentials each, the training set will be further split to implement $5$-fold cross-validation.
Note that this number of configurations is smaller compared to the $105.000$ potentials used in \cite{nelsonMachine2019}. 
We focused on this small-data regime due to the anticipated computational expense of generating data points with quantum algorithms.

\label{sec:potentials-and-ED}

\topic{Selecting the potentials}
To sample the random potential ${\pot}$ we follow a modified version of the approach in \cite{nelsonMachine2019}.
For each data point, we first sample a strength parameter ${W \in [0.005t,2.5t]}$ uniformly at random.
We then sample ${\pot}$ uniformly at random and calculate its standard deviation
${\sigma({\pot}) = \sqrt{\sum_j \mu_j^2 - (\sum_j \mu_j)^2}}$.
If the standard deviation ${\sigma({\pot}) < 0.4 t}$ we accept the potential and add it to our dataset, otherwise we reject and repeat the sampling procedure from the beginning.
This procedure produces a representative distribution of potential of varied strengths avoiding too large energy fluctuations, without requiring the solution of the Hubbard problem instance at this stage.

\topic{Exact diagonalization benchmark}
We construct an exact dataset to act as a baseline, solving all the problem instances (training and test set) with exact diagonalization.
For every problem instance $\hat{H}_{\pot}$, we construct the Hamiltonian matrix block corresponding to quarter filling and spin-singlet. 
We diagonalize it, obtaining the ground state $\ket{\psi_\gs({\pot})}$ and the energy $E({\pot})$. 
From the state we calculate the expected particle density vector $\dens_\gs({\pot})$.
We then compute the kinetic-interaction energy $\EUT({\pot}) = E({\pot}) - {\pot} \cdot \dens_\gs({\pot})$.
Thus, from each random configuration ${\pot}$, we obtain a pair $({\dens}_\gs, \EUT)$ representing the input and output of the exact Hubbard functional $\mathcal{F}$.

\subsection{Expectation value estimation} \label{sec:eve}

\topic{Sampling noise}
Assuming the exact ground state $\ket{\psi_\gs}$ of the Fermi-Hubbard Hamiltonian Eq.~\eqref{eq:hubbard_ham} can be prepared on a quantum device, measuring $\EUT$ and ${\dens_\gs}$ will still incur an error called \emph{sampling noise}.
To reproduce this effect we simulate expectation value estimation, by drawing samples from the measurement outcomes distribution defined by the Born rule on the ground state vector $\ket{\psi_\gs}$ produced by exact diagonalization.

\topic{Measurement scheme}
To define the measurements for expectation value estimation, we note that all the terms we want to measure are diagonal either in real space (density and Coulomb energy) or in Fourier space (kinetic energy) \cite{weckerProgress2015a}.
To estimate the real-space-diagonal terms, we can sample all the number operators $\hat{n}_{j, \sigma}$ at the same time.
Averaging $M$ samples we obtain estimates of each $\expval{\hat{n}_{j,\sigma}}$, and thus reconstruct an estimate $\tilde{\rho}_j$ for the density $\rho_j = \expval{\hat{n}_{j, \uparrow}} + \expval{\hat{n}_{j, \downarrow}}$.
The expectation value of the Coulomb energy
$
    \langle{\hat{U}}\rangle = u \sum_j \expval{\hat{n}_{j,\uparrow} \hat{n}_{j,\downarrow}}
$ 
can be estimated from the same samples by averaging the on-site correlators $\hat{n}_{j,\uparrow} \hat{n}_{j,\downarrow}$.
The kinetic energy operator $\hat{T}$ is diagonalized by expressing it in terms of the Fourier-space fermionic operators $\hat{c}_{k,\sigma} = 1/N \sum_{j=0}^{N-1} e^{i\,{j k 2 \pi}/N} \hat{c}_{j,\sigma}$, 
\begin{equation}
    \hat{T} = 2 t \sum_{k=0}^{N-1} \sum_{\sigma={\uparrow, \downarrow}} \cos(\frac{2 \pi k}{N}) \, \hat{n}_{k,\sigma} 
\end{equation}
where $\hat{n}_{k,\sigma} = \hat{c}^\dag_{k,\sigma} \hat{c}_{k,\sigma}$.
The operators $\hat{n}_{k,\sigma}$ can be sampled at the same time; another $M$ samples are taken and averaged to obtain $\expval{\hat{n}_{k,\sigma}}$, and their results combined to estimate $\langle{\hat{T}}\rangle$.
The estimate $\tildeEUT_\text{EVE}$ of $\EUT$ is finally calculated by summing the kinetic and Coulomb contributions.
For the central limit theorem the estimate of $\tilde{F}$ is unbiased and asymptotically normal, with variance proportional to $1/M$.

\topic{Implementation on qubits}
Assuming we have a qubit-based quantum device and Jordan-Wigner Fermion-to-qubit encoding \cite{jordanUeber1928,whitfieldSimulation2011}, the real-space samples are simply obtained by measuring all qubits in the computational basis as $\hat{n}_{j,\sigma} = (1-Z_{j,\sigma})/2$ (where $Z_{j,\sigma}$ is the Pauli Z-operator on the qubit that encodes the spin-orbital ${j,\sigma}$).
In order to sample the plane wave number operators $\hat{n}_{k,\sigma}$, we need to perform a basis rotation circuit implementing the fast fermionic Fourier transform \cite{babbushLowDepth2018}: a circuit of depth $O(N)$ acting on the two spin sectors independently.

\subsection{Variational quantum eigensolver}
\label{sec:vqe}

\topic{definiton of VQE}
The VQE is a hybrid quantum-classical method that combines a quantum subroutine with a classical optimizer, aiming to estimate the ground state energy of a Hamiltonian ~\cite{peruzzoVariational2014}. 
The quantum subroutine prepares on a quantum device an ansatz state $\ket{\psi(\boldsymbol{\theta})} = U(\boldsymbol{\theta}) \ket{\psi_{\gs}}$ dependent on a set of classical parameters $\boldsymbol\theta$; and measures its expected energy $E(\boldsymbol{\theta}) = \expval{H}{\psi(\boldsymbol{\theta})}$ through EVE.
This energy is then minimized over the parameters $\boldsymbol\theta$ using a classical optimizer which calls the quantum subroutine.

\topic{VQE errors (expressibility and optimization error)}
The VQE is a heuristic method, which means that convergence to the true ground state energy is not guaranteed~\cite{tillyVariational2022}.
The manifold of states $\psi(\boldsymbol{\theta})$ that can be represented by the chosen ansatz is, in general, much smaller than the Hilbert space of the problem.
This is a feature of VQE: restricting the number of parameters is the only way to make the problem feasible for the classical optimizer.
The minimum energy ansatz state will thus only approximate the ground state, with its energy $\min_{\boldsymbol{\theta}}E(\boldsymbol{\theta})$ guaranteed to be larger than the ground state energy by the variational principle.
The quality of the results depends on the choice of ansatz, the performance of the classical optimizer, and is further affected by hardware noise and sampling noise.
In this work, we focus on isolating the effect of the errors intrinsic to the VQE due to ansatz expressibility and optimization. 
We use a classical state-vector simulator to extract expectation values without sampling noise, which is studied separately through EVE.
Both the energy and the electronic density of the optimal state are affected by the VQE errors.

\topic{Ansatz in VQE}
Variational ansatz families fall into two main categories: hardware-efficient and physically inspired. Physically inspired ansätze, like the Unitary Coupled Cluster (UCC)~\cite{taubeNew2006, peruzzoVariational2014} consider essential system properties but often require high circuit depths, making their implementation challenging for near-term quantum devices. 
Hardware-efficient ans{\"a}tze instead maximize the number of parameters per circuit layer, but may still require many layers to approximate the ground state and do not take into consideration the problem symmetries.
We will consider two ans\"{a}tze in this work: the well-known Variational Hamiltonian Ansatz (VHA)~\cite{weckerProgress2015a} and the Number Preserving Fabric (NPF)~\cite{anselmettiLocal2021} ansatz, which combine the advantages of both categories.

\topic{VHA}
The VHA is one of the earliest proposed ans{\"a}tze for VQE study of the Fermi-Hubbard model \cite{weckerProgress2015a,weckerSolving2015}, and has found success in hardware experiments \cite{cadeStrategies2020, stanisicObserving2022}. 
This ansatz is inspired by the adiabatic algorithm~\cite{farhiQuantum2000, farhiQuantum2001} and formally equivalent to the quantum alternating operator ansatz (QAOA) \cite{farhiQuantum2014,hadfieldQuantum2019}.
It distinguishes itself from the usual VQE ans{\"a}tze by incorporating the values of $\pot$ defining the problem instance into the ansatz defintion.
The circuit structure for the Hubbard model, for parameters $\boldsymbol{\theta}=\left\{\theta_{i,j}\right\}$, is
\begin{align}\label{eq:vha}
    \ket{\psi\left\{\theta_{i,j}\right\}} = \prod_{i=1}^p \left[ e^{-i \hat{T}_e \theta_{i,0}} e^{-i \hat{T}_o \theta_{i,1}} e^{-i \hat{V}_{\pot} \theta_{i, 2}} e^{-i \hat{U} \theta_{i, 3}}   \right] \ket{\psi_0},
\end{align}
thus able to represent a first order trotter decomposition in $p$ slices of the adiabatic evolution to $H$ from $H^0$, commonly taken as the non-interacting Hamiltonian of which $\ket{\psi_0}$ is the ground state. 
In Eq.~\eqref{eq:vha}, $\hat{T}_e$ ($\hat{T}_o$) are all hopping terms on even (odd) sites in Eq.~\eqref{eq:kinetic_term}.

\topic{NPF}
The NPF ansatz takes another strategy which is, in contrast to VHA, independent of the problem. 
It composes the two fundamental interactions contained in a spin-preserving two-body Hamiltonian; a spatial orbital rotation and a double excitation. 
The parameterized composition of these operations we call $Q(\theta, \phi)$, and they are arranged in a brick-wall pattern to maximize the gate density, resulting an expressive ansatz with only local gates. For the exact form of the ansatz, see Ref.~\cite{anselmettiLocal2021}.

\topic{Comparison of VHA vs NPF (nr. parameters and optimizer)}
While both the VHA and NPF ansatz are physically inspired and preserve key symmetries like total spin and number of particles, we highlight here some differences between the two. The number of parameters in the VHA equals $4$ per layer by inspection of Eq.~\eqref{eq:vha}.
The NPF ansatz has two parameters per $Q$-block, of which there are $\frac{N_q}{2}-1$ per layer. In our case $N_q = 2N = 16$, resulting $14$ parameters per layer. Although the NPF ansatz has more parameters per layer, its higher gate density means it does not necessarily result in deeper circuits.
VQE cost functions are often highly irregular and non-convex, with numerous local minima. Hence, the choice of classical optimizer is crucial~\cite{bonet-monroigPerformance2023}. The VHA, being a more structured ansatz, may have a more irregular cost landscape, making it prone to getting stuck in local minima. To address this, we use the gradient-free Constrained Optimization by Linear Approximation (COBYLA) optimizer. For the NPF, the abundance of parameters offers more flexibility but less structure, and empirically, following the gradient in this cost landscape tends to work better. Therefore, we employ the Sequential Least Squares Programming (SLSQP) optimizer for the NPF ansatz.

\section{Learning the functional}\label{sec:learning_functional}

In this section we present the chosen machine learning models and the details of the training procedure.
Alongside the main neural-network model, for which we provide a detailed result analysis in Sec.~\ref{sec:results}A, we present a comparative study of a set of out-of-the box models in Sec.~\ref{sec:model-selection}.

\subsection{Method} \label{sec:method}

\topic{CNN model}
The main ML model we consider is an adaptation of the convolutional neural network (CNN) model proposed by Nelson \emph{et al.} \cite{nelsonMachine2019}.
The model consists of a sequence of one 1-dimensional periodic convolutional layer with 8 local features and kernel size 3, two dense layers with 128 features each, and a final dense layer with a single output.

Before being processed by the CNN, the inputs $\dens_X$ are normalized.
Each element $\rho_{j}$ of the density $\dens$ is normalized using the mean $\bar\rho$ and standard deviation $\sigma_\rho$ of the whole training set:
\begin{equation}
    \rho_j \mapsto (\rho_j - \bar\rho)/\sigma_\rho =: \text{input}_j,
\end{equation}
(where we dropped the method index $X$ for convenience of notation).
Conversely, the output of the model is rescaled by the standard deviation of the training energies $\sigma_{\EUT}$ and shifted by their mean $\bar{\EUT}$:
\begin{equation}
    \text{output} \mapsto \sigma_{\EUT} \cdot \text{output} + \bar{\EUT} := \mathcal{F}[\dens].
\end{equation}

\topic{Cross-validation}
In potential real-world application of our method to noisy data from systems too large for exact benchmarking, the training data becomes the sole source of information for model evaluation. Thus, model selection, hyperparameter optimization and validation must be conducted exclusively using the available (noisy) dataset.
To address this, we use $5$-fold cross-validation \cite{kohaviStudy1995}.
This involves partitioning the noisy training dataset into five equally sized subsets (folds). 
During each iteration, one fold is set aside as the validation set, while the remaining four folds are used to train the model.
The performance metrics obtained from each fold are then averaged to provide a comprehensive evaluation of the model performance.

\topic{Symmetry augmentation}
The target functional $\mathcal{F}[\dens]$ is invariant under translations, mirror symmetry, and their combination.
This means that $\mathcal{F}[\dens]$ does not change under the transformation
\begin{equation}
    \rho_j \mapsto \rho_{(\pm j + k) \operatorname{mod} N}
    \,,\quad \forall k \in \{0, ..., N-1\}.
\end{equation}
We exploit this to augment the training dataset: from each pair $(\tilde{\dens}_X, \tilde{\EUT}_X)$ we construct 16 data points, applying the $N=8$ shifts and $N=8$ mirror-and-shifts to the density and copying the energy \cite{nelsonMachine2019}. 
This data augmentation is performed after the cross-validation split, on the training and validation splits separately. 
The data points within the training split are then scrambled before dividing in batches for training.

\topic{optimization/Early stopping/keras}
The model is trained with the Adam optimizer \cite{kingmaAdam2017}.
To avoid overfitting, we implement an early stopping strategy which monitors the validation loss during training, halting when this loss stops decreasing.
The model and its training are implemented using the TensorFlow-Keras library \cite{abadiTensorFlow2016,chollet2015keras}.

\subsection{Alternative model selection}
\label{sec:model-selection}

\begin{figure}[t]
    \centering
    \includegraphics[width=\columnwidth]{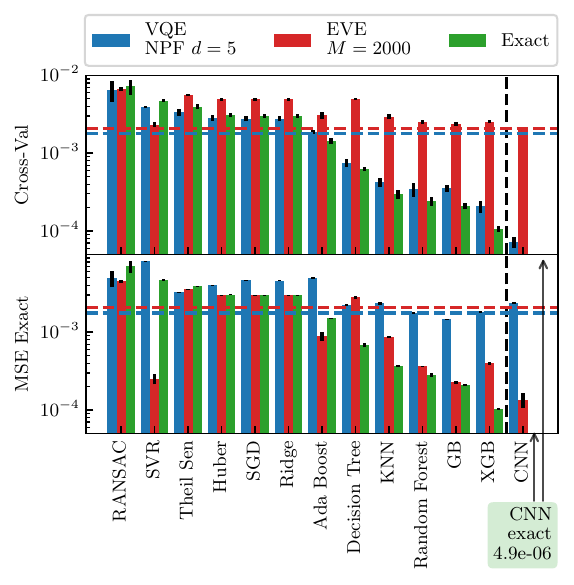}
    \caption{\textbf{Model Selection:}
    Comparison of the performance of various regression models, trained and validated using an variational quantum eigensolver (VQE) (Number Preserving Fabric Ansatz with depth $d=5$), expectation value estimation (EVE) with $M=2000$ shots, and exact data. 
    The top panel displays the performance based on $5$-fold cross validation MSE, computed by comparing the predictions made by models trained on individual folds against the (noisy) data across their respective validation sets.
    The bottom panel benchmarks the model performance by evaluation on exact densities from a test dataset and comparing them with the corresponding exact energies.
    Dashed lines indicate the baseline MSE of the noisy datasets used for training and validation against the exact data.
    Based on the cross-validation scores in the top panel, the CNN model from Ref. \cite{nelsonMachine2019} outperforms all the out-of-the-box regression models we tested for the learning task at hand. 
    The benchmarking results in the bottom panel indicate that the learned representation extrapolates well to noiseless data, achieving consistently low errors across the board. 
    This suggests that among the presented models, the CNN has the strongest ability to capture the underlying functional relationship between density and energy, even when trained on noisy data.
    }
    \label{fig:model_selection}
\end{figure}

\topic{model selection}
Given the changes in the underlying statistical structure of the data in this work compared to Ref.~\cite{nelsonMachine2019}, we evaluate the suitability of the CNN model by comparing its performance with other machine learning models known for their robust regression capabilities and ability to handle noise and outliers. 
The objective of this section is to determine if any of these alternative models can enhance the learning task with our noisy data.
Our findings indicate that the CNN model is the best performing model.
Nonetheless, we present our comparative study for completeness.

\topic{models}
The ensemble of models to be evaluated comprises $11$ regression models, 
including the convolutional neural network (CNN) model from Ref.~\cite{nelsonMachine2019}, three linear regression-based approaches known for their outlier robustness (Huber regression ~\cite{owenRobustHybridLasso2007}, Theil-Sen regression (TS)~\cite{dangTheilSenEstimatorsMultiple}, and random sample consensus (RANSAC) regression~\cite{fischlerRandom1981,cantzlerRandomSampleConsensus}), five decision tree models (AdaBoost (AB)~\cite{freundShortIntroductionBoosting}, random forest (RF)~\cite{breimanRandom2001}, gradient boosting (GB)~\cite{natekinGradient2013}), support vector regression (SVR)~\cite{druckerSupport1996}, nearest neighbor regression (KNN)~\cite{altmanIntroduction1992a}, and XGBoost (XGB)~\cite{chenXGBoost2016}.
For the CNN model, we adopted the hyperparameters from Nelson \emph{et al.} \cite{nelsonMachine2019}, while for the other models, we used standard baseline hyperparameters.

\topic{x-validation}
To evaluate model performance, we present results for models trained on data generated from both the VQE (using an NPF ansatz with depth $d=5$) and EVE ($M=2000$ shots) methods. 
These two instances are chosen to represent the typical characteristics of VQE and EVE datasets, and we focus exclusively on them for clarity and readability.
For benchmarking purposes, we also include performance metrics for models trained on exact data. 
In scenarios where our method is applied to noisy data from systems that are too large for exact benchmarking, the cross-validation of training data will be the only resources available. 
Consequently, model selection and hyperparameter optimization must be based solely on these datasets:
For this, we employ $5$-fold cross-validation, as introduced in Sec. \ref{sec:method}. 

\topic{plot overview}
The comparative performance of the models is illustrated by the mean and standard deviation of their cross-validation scores, as shown in the top panel of Fig. \ref{fig:model_selection}.
Additionally, both panels display the baseline mean-squared error (MSE) of the noisy data used for training and validation, indicated by dashed lines.
The bottom panel of Fig. \ref{fig:model_selection} illustrates the models' performance compared to the exact benchmark on a separate test set, generated with a distinct set of potentials not used during training.
While this information is not directly used in the model selection process, it provides insights into the models' ability to generalize to noiseless data. 

\topic{plot interpretation}
The cross-validation scores for models trained on EVE data never fall below the baseline error of the training data. 
This is because the validation dataset is also noisy, and with EVE's noise being randomly distributed with a zero mean, the model cannot effectively learn or predict the noise pattern.
Among the models tested, those achieving the lowest validation scores on this dataset, aside from the optimized CNN model, are Support Vector Regression (SVR), Gradient Boosting (GB), Random Forest, and XGBoost (XGB). 
These models also perform the best in predicting noiseless data.
We observe the characteristic performance that a cross-validation score close to the baseline of unbiased noise indicates that the underlying model has learned patterns that extrapolate to some extent to noiseless data. 

The performance of models trained on VQE data differs noticeably when tested against exact data. 
As discussed in Section~\ref{sec:vqe}, the optimization error inherent in our VQE data can lead to an overestimation of the ground state energy for a given model instance. 
This results in biased noise that affects the training process. 
Consequently, the model inadvertently learns to fit this biased noise, which does not exist in the exact data, leading to a worse performance when testing the generalizability. 
When the MSE between the VQE-generated energies and the exact energies (indicated by the dashed blue line) is less than or equal to the predicted MSE from cross-validation, it suggests that the model has absorbed the bias present in the VQE data. 
Specifically, for models 7-13 trained on VQE data, the validation MSE is smaller than the baseline error of the training data.
This observation implies that (1) there is an (expected) bias in the error of VQE and (2) these models learn to predict this bias.

\topic{CNN FTW}
Based on the cross validation data, the CNN model from Ref. \cite{nelsonMachine2019} performs best on all datasets.
While gradient boosting performs slightly better on the presented VQE dataset (NPF ansatz with $d=5$), general performance across VQE datasets is at best comparable to the CNN model.
We thus continue further analysis with the CNN model only.

\section{Results}\label{sec:results}

\begin{figure}[t]
    \centering
    \includegraphics[width=\columnwidth]{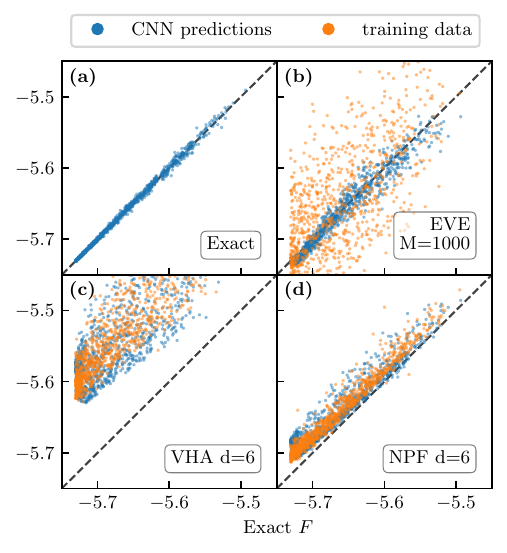}
    \caption{
    \textbf{Noisy data and model predictions:}
    Comparison of exact ground state kinetic-interaction energies $\EUT$ with training data and model predictions across four different methods for generating density-energy pairs.
    (a) Illustrates  the model's ability to learn a functional from exact data, accurately capturing the relationship between density and ground state energies obtained from exact diagonalization.
    (b-d) Demonstrates the model's capacity to learn a functional which generalizes to the exact data, resulting in predictions that are on average closer to the ground truth than the raw data.
    These panels include:
    expectation value estimation (EVE) with $M=1000$ shots and
    variational quantum eigensolver (VQE) with Variational Hamiltonian (VHA) / Number-Preserving Fabric (NPF) Ansatz  of depth $d=6$.}
    \label{fig:CNN}
\end{figure}

In this section, we benchmark the CNN model trained on noisy datasets, analyzing its performance as a function of both the type and magnitude of noise in the training data.
We test the models on two tasks: energy prediction, discussed in Sec.~\ref{sec:energy-prediction-results}, and density optimization, discussed in Sec.~\ref{sec:ks-results}.
All benchmark are conducted using the same test set, consisting of 1,000 new problem instances with random potentials $\pot_\text{test}$.
These are solved with exact diagonalization obtaining a set of exact test densities $\dens_\text{test}$ and energies $\EUT_\text{test}$.

\subsection{Energy prediction}
\label{sec:energy-prediction-results}

In the energy prediction task, we assess the model's accuracy by comparing its predictions on the exact densities, $\mathcal{F}^\text{ML}[\dens_\text{test}]$, against the exact energies $\EUT_\text{test}$.
Figure \ref{fig:CNN} summarizes this comparison for four models, each trained on a datasets constructed with a different method: exact diagonalization, EVE $(M=1000)$, VHA-VQE $(d=6)$, and NPF-VQE $(d=6)$.
The figure also reports the noisy training energies $\EUT$, compared to the respective exact value.
The model trained on exact data serves as a benchmark, representing an upper limit of performance.
Models trained on the same amount of noisy data points are expected to perform less accurately.

The EVE data are spread uniformly around the diagonal line, as sampling noise is unbiased.
The model trained on this data demonstrates improved energy predictions, which cluster closer to the diagonal line compared to the noisy training data.

Conversely, the VQE data suffer from a positive bias due to the limited expressive power of the ansatz.
Variational methods always approximate the ground state energy from above, $\tilde{E}_\text{VQE} \geq E$. 
As a consequence, $\tilde{F}_\text{VQE}$ typically (albeit not always) overestimates $\EUT$.
We observe this bias is larger for smaller values of $\EUT$, particularly for the $d=6$ NPF ansatz.
This effect is likely due to the variation in the strength of $\pot$, which influences the efficiency of VQE -- larger potentials cause stronger localization, thereby increasing kinetic and interaction energy as electrons are more confined. 
Localization also reduces ground state correlations, which can enhance the performance of limited-depth VQEs.
The predictions of the models trained on VQE data are distributed similarly to their respective training data, indicating that the model learns the inherent bias.

\begin{figure}[t]
    \centering
    \includegraphics{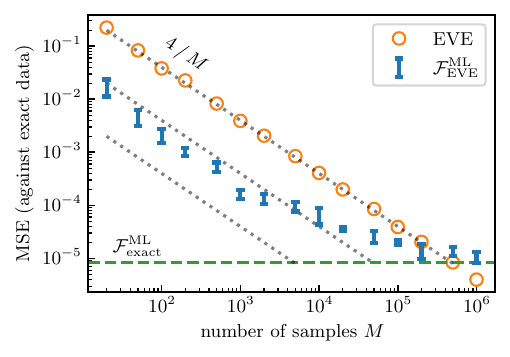}
    \caption{
    \textbf{Learning improvement for sampling noise:}
    Scaling of the Mean Squared Error (MSE) on $\EUT$ as a function of $M$, the number of sampling shots used in the data generation for the training set.
    The errors are evaluated against exact diagonalization data.
    The orange circles represent the MSE of the raw EVE data, demonstrating how the estimate variance decreases proportionally to $1/M$.
    The top dotted black line indicate this scaling, the others indicate a MSE reduction by one and two orders of magnitude.
    The dashed green line marks the baseline of the model trained and evaluated on exact data, providing a reference point for assessing the relative performance of the models.
    Blue markers show the MSE of the CNN model trained with EVE data, evaluated on exact densities from a test set and compared to the corresponding energies.
    The error bars indicate the standard deviation over the 5 cross-validation folds.
    The model predictions improve on the raw EVE error by more than an order of magnitude in the high-noise region, flattening to the baseline provided by the model trained on noiseless data when the noise becomes small enough.
    }
    \label{fig:eve-result}
\end{figure}

\topic{MSE}
To systematically compare models trained on datasets with a varying noise strengths, we summarize the results of the energy prediction task with the mean squared error (MSE)
\begin{equation}
    \operatorname{MSE}_{\mathcal{F}^\text{ML}} = 
    \frac{1}{\text{test size}} \sum_\text{test set} |\mathcal{F}^\text{ML}[\dens_\text{test}] - \EUT_\text{test}|^2.
\end{equation}
This is compared to the raw MSE of the method used to generate the training data, calculated directly from the training set as 
\begin{equation}
    \operatorname{MSE}_X = \frac{1}{\text{train size}} \sum_\text{train set} |\tildeEUT_X - \EUT|^2.
\end{equation}

\topic{EVE-trained models}
Figure \ref{fig:eve-result} shows the performance of the model trained on EVE data as a function of the number of samples $M$.
The squared error in the training data decreases proportionally to $1/M$, consistent with the standard sampling limit.
(The factor 4 arises from the average ground-state variance of the two operators $\hat{T}$ and $\hat{U}$, which are sampled separately as discussed in Sec.~\ref{sec:eve}.)
The model's predictions improve on this MSE by up to factor 20 in the case of large sampling noise.
As the sampling noise decreases, the model's performance saturates to that of the model trained on exact data.
It is important to note that the total number of samples needed to generate the training dataset is $1000\,M$, uniformly distributed over the $1000$ different problem instances.
The improvement is due to the model's ability to synthesize the information from all these training instances, distilling the underlying trend while filtering out the unbiased noise.

\begin{figure}[t]
    \centering
    \includegraphics{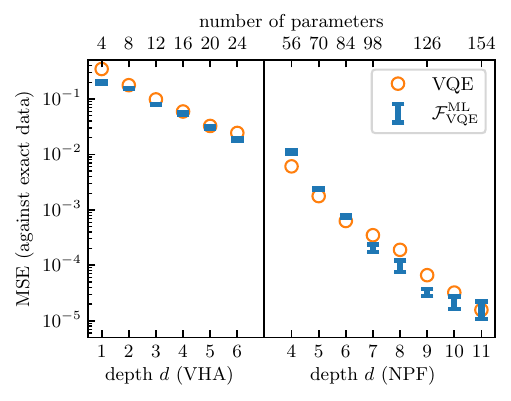}
    \caption{
    \textbf{Models learn the bias from VQE data:}
    Scaling of the Mean Squared Error (MSE) for different models evaluated against exact data as a function of $d$, the depth of the circuit used in the data generation for the training set, and the type of Ansatz VHA/NPF.
    The orange line represent the MSE of the raw VQE data, demonstrating how the error decreases as $d$ increases. 
    Red markers show the mean of 5-fold cross-validation performance (black bars represent the individual model performance) of machine learning models trained to predict VQE data and evaluated against exact data, highlighting some improvements against the raw data. 
    }
    \label{fig:vqe-result}
\end{figure}

\topic{VQE-trained models}
Figure \ref{fig:vqe-result} presents the same performance benchmark for the model trained on VQE data.
The MSE of the VQE, for both ans{\"a}tze considered, decreases exponentially with the depth of the ansatz.
Here, the depth $d$ represents the number of repeated layers in the ansatz, while the upper-axis labels indicate the number of parameters.
For depths larger than $d=6$ the optimization of the VHA becomes significantly harder, thus we explore the larger-parameter-number regime with the less-structured NPF ansatz.
The learned functional generalizes the training dataset successfully, showing a MSE comparable to the respective training data for all depths and both ans{\"a}tze.
Unlike the models trained on EVE, those trained on VQE data do not manage to learn the underlying functional while filtering out noise. 
This limitation is due to the bias intrinsic to the VQE errors, which is learned by the model.
Although there is some quasi-stochastic error in the VQE results, caused by the optimization converging to different local minima of the energy, the dominant factor appears to be the biased error due to the limited expressibility of the ansatz.

\subsection{Density optimization}
\label{sec:ks-results}

\begin{figure*}[t]
    \centering
    \includegraphics{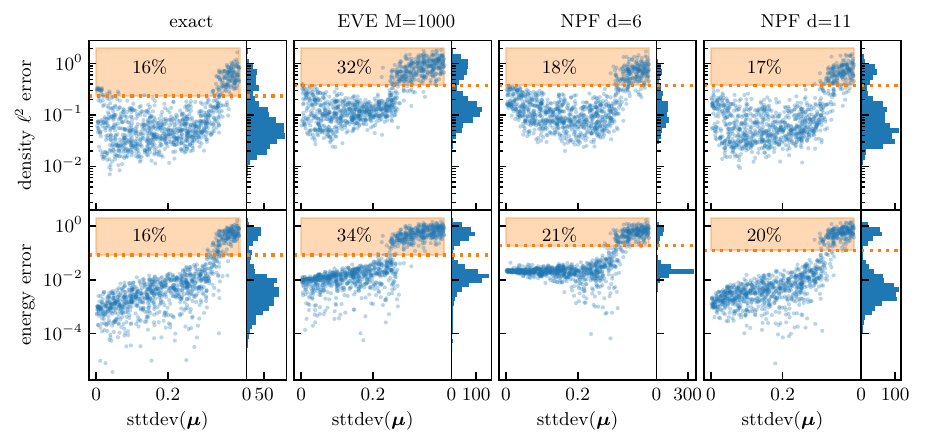}
    \caption{
        \textbf{Application of the trained model to density optimization:}
        Results of Kohn-Sham-like density optimization on a test set of 1000 new instances defined by random potentials, with the Hubbard functional defined by the CNN trained on various datasets.
        The performance is characterized by the density error in $\ell^2$ norm $\lVert \dens_\text{opt} - \dens_\text{exact}\rVert_2$ (top row) and by the absolute error on the kinetic-interaction energy $|f[\dens_\text{opt}] - \EUT_\text{exact}|$ (bottom row).
        The scatter plots show the error dependence on the standard deviation of the potential $\lVert {\pot} - \bar{\mu} \rVert_2$, and the histograms summarize the errors.
        The histograms show the errors tend to follow a bimodal distribution; the percentage of points in the higher-error mode of the distribution (shaded area above the dotted line) is annotated on each plot.
    }
    \label{fig:density-optimization}
\end{figure*}

\topic{Motivation}
The main application of DFT functionals lays in solving new problem instances through density optimization.
The goal of density optimization is equivalent to usual Kohn-Sham self-consistent field, but rather than optimizing a set of single-particle orbitals together with an exchange-correlation functional, we directly optimize the total energy functional in Eq.~\eqref{eq:total-energy-functional} with respect to the density $\dens$ (subject to appropriate constraints).
In this section we explore the performance in this application of the ML functional learned from noisy data using the CNN model.

\topic{Task description}
For each test set potential $\pot_\text{test}$ and each considered learned functional $F^\text{ML}$, we construct the total energy functional
\begin{equation} \label{eq:total-energy-functional-ml}
    \mathcal{E}^\text{ML}[\dens] = \mathcal{F}^\text{ML}[\dens] + \pot_\text{test} \cdot \dens.
\end{equation}
We then minimize $\mathcal{E}^\text{ML}[\dens]$ under the constraints of $\rho_j \in [0, 2]$ for each site $j$ (positive and bounded occupation) and $\sum_j \rho_j = 4$ (fixed total number of particles); this yields predictions for the ground state density $\dens^*_\gs$ and energy $E^* = \mathcal{E}[\dens^*]$.
The minimization is performed using the Sequential Least Squares Programming (SLSQP) method implemented in the SciPy package \cite{virtanenSciPy2020}, exploiting automatic differentiation of the CNN model to obtain analytical gradients of $\mathcal{E}[\dens]$.

\topic{Description of Figure \ref{fig:density-optimization} }
Figure \ref{fig:density-optimization} shows the resulting errors, for a selection of models trained on datasets constructed with different methods with different accuracy: exact diagonalization, EVE $(M=1000)$, NPF-VQE $(d=6)$, and NPF-VQE $(d=11)$.
For energy we consider the absolute deviation $|E^* - E_\text{test}|$, and for density we consider the error in $\ell^2$-norm \begin{equation}
    \lVert \dens^* - \dens_{\text{test}}\rVert_2
    =
    \sqrt{\sum\nolimits_j |\rho^*_j - \rho_{\text{test}\,j}|^2}.
\end{equation}
In the scatter plots we show these errors for each of the test set instances versus the strength of the potential $\pot$ ---
as the average value of the potential $\bar\mu = \sum_j \mu_j$ shifts the functional Eq.~\eqref{eq:total-energy-functional-ml} by a constant and does not affect optimization, the strength of the potential is characterized as its standard deviation $\operatorname{sttdev}(\pot) = \lVert \pot - \bar\mu \rVert_2$.
To the right of each scatter plot, the distribution of the errors in summarized in a histogram.
As all histograms display bimodal distributions, with two clearly separated lobes at lower- and higher-errors.
On each histogram, we show the line that separates the two modes, and report the fraction of training data in the higher-error mode.

\topic{Interpretation of Figure \ref{fig:density-optimization}}
The results in the left panel of Fig.~\ref{fig:density-optimization} show that, even when using the model trained on exact data, density optimization does not always converge to the correct result.
In particular, for the largest potential strengths the optimization yields densities and and energies with a large error.
We attribute this to the limited availability of training points near the edges of the potential distribution.
In the region of smaller potential strengths, at the center of the distribution of $\pot$, has more data available; there the model can learn and generalize correctly.
Farther from the center of the distribution, where the potential strength is larger, few data points are available for a wider area of possible cases to cover.
We estimate $16\%$ of the test set potentials are in this edge area; and a similar fraction of training points are available there.
While this unavoidable for a continuous and unbounded distribution of problem instances, one can mitigate the problem by training the model on a set of potentials wider than the application range.

Conversely, for the smallest potentials, the density optimization sometimes yields correct energies and wrong densities.
This occurs because, for very small potentials, the considered Hubbard model can exhibit two quasi-degenerate states with similar energies but significantly different densities.
As a consequence $\mathcal{E}[\dens]$ has multiple local minima for very distant densities, and the local optimizer we use sometimes finds a wrong density with a good associated energy.

Both of these effects, for the smallest and largest potential strengths, affect the results of the models trained using noisy data in a qualitatively similar way.
Noisy training data causes an overall increase of error on the densities and energies, as well as shrinks the range of potentials on which the results of energy optimization are reliable.
In the case of VQE with lower depth (NPF $d=6$), we can see a concentration of the energies due to the large positive bias of the training data, which dominates the energy error.

\section{Discussion and outlook}\label{sec:conclusion}

\topic{conclusion}
In this work, we demonstrated the application of machine learning models motivated by density functional theory to generalize and enhance a set of data subject to noise characteristic of near-term quantum algorithms.
We showed that meaningful density functionals can be learned from a small amount of noisy training data and benchmarked the performance of these models in their usual DFT applications.
Comparing the performance of the learned models to the respective training data, we showed a significant improvement is only achieved when the noise in the input data is unbiased (Fig.~\ref{fig:eve-result}).
This implies that learning techniques are well-suited to improve results from quantum algorithms that suffer from sampling noise.
However, even when the trained models learn the inherent dataset bias, the proposed technique could be useful to generalize the dataset to new problem instances (Fig.~\ref{fig:density-optimization}).
In the near future, quantum devices might be able to generate approximate solutions for quantum simulation problems that are beyond the reach of classical computation but remain too noisy to be directly useful.
Nonetheless, this data could be valuable for supplementing the training of classical machine learning models, possibly in combination with other classically-generated data.
In this context, it is important to characterize the effect of noise in quantum data on the trained models.

\topic{Joint noise study}
Our study focused on analyzing the (biased) VQE error and (unbiased) sampling noise separately. 
However, in practical implementations on quantum hardware, both sampling and optimization noise critically influence the performance and accuracy of VQE algorithms. 
Additionally, hardware noise is a significant factor. 
Investigating the combined effects of these noise sources presents a promising direction for future research.

\topic{Scaling with number of samples and system size}
Furthermore, we concentrated on learning from a very limited dataset, comprising only 1,000 training and validation instances. 
To better understand the trade-offs involved in achieving functionals with useful precision, it would be relevant to study the performance of learning as a function of both dataset size and problem size.
Additionally, advancing towards realistic applications will require the development of techniques that enable learning from mixed datasets, combining results from approximate classical algorithms and quantum data. 
In this context, transfer learning might offer valuable insights ~\cite{panSurvey2010}.

\topic{Which functional to learn}
In this study, we focused on learning Hohenberg-Kohn density functionals, a cornerstone of density functional theory with extensive literature support; however, the choice of learning targets can generally vary depending on the specific application.
In the future, we envision betond-classical quantum computations contributing valuable data to further enhance a wider range of existing machine learning-based functionals.
Another potential learning target is the exchange-correlation functional used in Kohn-Sham density functional theory, which differs from our functional by a classical estimate of the kinetic energy expectation value, based on a product state. 
Alternatively, learning targets could be drawn from 1-particle reduced density matrix functional theory, where the underlying compressed representation of the quantum state is its 1-particle reduced density matrix ~\cite{pernalReduced2016}.
Finally, one could consider directly learning the mapping $\pot \to E$, also known as the Hohenberg-Kohn mapping \cite{brockherdeBypassing2017}. 
A comparative study could help determine which of these learning targets is most resilient to noise in the data, offering better adaptability in real-world scenarios.

\section*{acknowledgements}
We thank Vedran Dunjko and Patrick Emonts for their useful feedback.
This work was supported by the Dutch National Growth Fund (NGF), as part of the Quantum Delta NL programme.
S.P.~and E.K.~acknowledge support from Shell Global Solutions BV.
J.T.~acknowledges the support received from
the European Union’s Horizon Europe research and innovation programme through the ERC StG FINE-TEA-SQUAD (Grant No.~101040729).
This publication is part of the `Quantum Inspire - the Dutch Quantum Computer in the Cloud' project (with project number NWA.1292.19.194]) of the
NWA research program `Research on Routes by Consortia (ORC)', which is funded by the Netherlands Organization for Scientific Research (NWO).
The views and opinions expressed here are solely those
of the authors and do not necessarily reflect those of the
funding institutions.
Neither of the funding institutions
can be held responsible for them.

\section*{Code and Data Availability}
The density-energy pair data used for training the machine learning models, along with the code used to generate it and to reproduce the plots in this paper, is available at the repository: \href{https://github.com/StefanoPolla/DFTQML}{https://github.com/StefanoPolla/DFTQML}.

\section*{Competing Interests}
The authors declare no competing interests.

\section*{Contributions}
E.P., E.K.~and S.P.~conceived the project and developed a first prototype of the code.
S.P.~developed the codebase for data generation and learning.
E.K.~developed and optimized VQE.
F.F.~performed the model selection studies.
E.K., F.F.~and S.P.~wrote the manuscript, which all authors contributed to reviewing.
J.T., E.vN.~and S.P.~supervised the work and provided funding.

\bibliography{zotero-export, manual-refs}

\end{document}